\documentstyle[12pt]{article}
\newcommand{\ga}{$g_{\pi NN}(q^2)$}
\newcommand{\la}{$\Lambda_{\pi NN}$}

\begin{document}

\begin{titlepage}

\begin{flushright}
       {\bf UK/94-01}  \\
       {\bf hep-lat/9406007} \\
       June 1994      \\
\end{flushright}
\begin{center}

{\bf {\LARGE $\pi NN$ and Pseudoscalar Form Factors \\
\vspace{1ex}
from Lattice QCD}}

\vspace{1.5cm}

{\bf  K.F. Liu, S.J. Dong, and Terrence Draper} \\ [0.5em]
{\it  Dept. of Physics and Astronomy  \\
      Univ. of Kentucky, Lexington, KY 40506} \\ [1em]
{\bf  Walter Wilcox}\\ [0.5em]
{\it  Dept. of Physics, Baylor Univ., Waco, TX 76798}

\end{center}

\vspace{1cm}

\begin{abstract}

The $\pi NN$ form factor $g_{\pi NN}(q^2)$ is obtained from a quenched
lattice QCD calculation of the pseudoscalar form factor $g_P(q^2)$ of
the proton with pion pole dominance. We find that $g_{\pi NN}(q^2)$
fitted with the monopole form agrees well with the Goldberger-Treiman
relation and is much preferred over the dipole form. The monopole mass
is determined to be $0.75 \pm 0.14\,{\rm GeV}$ which shows that $g_{\pi
NN}(q^2)$ is rather soft. The extrapolated $\pi N$ coupling constant
$g_{\pi NN} = 12.7 \pm 2.4$ is quite consistent with the
phenomenological values. We also compare $g_{\pi NN}(q^2)$ with the
axial form factor $g_A(q^2)$ to check the pion dominance in the induced
pseudoscalar form factor $h_A(q^2)$ vis \`{a} vis chiral Ward identity.

\vskip\baselineskip
\noindent
PACS numbers: 12.38.Gc, 14.20.Dh, 13.75.Gx, 13.60.-r

\end{abstract}

\vfill

\end{titlepage}

The $\pi NN $ form factor $g_{\pi NN}(q^2)$ is a fundamental quantity
in low-energy pion-nucleon and nucleon-nucleon dynamics. Many dynamical
issues like $\pi N$ elastic and inelastic scattering, $NN$ potential,
three-body force (triton and ${}^{3}{\rm He}$ binding energies), pion
photoproduction and electroproduction all depend on it. Similarly,
the pseudoscalar form factor is important in testing low-energy
theorems, the chiral Ward identity and the understanding of the
explicit breaking of the chiral symmetry. Yet, compared with the
electromagnetic form factors and the isovector axial form factor of the
nucleon, the pseudoscalar form factor $g_P(q^2)$ and the $\pi NN$ form
factor $g_{\pi NN}(q^2)$ are poorly known either experimentally or
theoretically. 

Notwithstanding decades of interest and numerous work, the shape and
slope of $g_{\pi NN}(q^2)$ remain illusive and unsettled. Upon
parametrizing \ga\,\, in the monopole form 
\begin{equation} \label{pnnff}
g_{\pi NN}(q^2) = g_{\pi NN} \frac{\Lambda_{\pi NN}^2 - m_{\pi}^2}
{\Lambda_{\pi
NN}^2 - q^2}
\end{equation}
with $g_{\pi NN} \equiv g_{\pi NN}(m_{\pi}^2)$, the uncertainty in the
parametrized monopole mass $\Lambda_{\pi NN}$ can be as large as a
factor 2 or 3. For the sake of having a sufficiently strong tensor
force to reproduce the asymptotic D- to S- wave ratio and the
quadrupole moment in the deuteron,  \la\, is shown to be greater than
$1\,{\rm GeV}$~\cite{er83}. Consequently, \la\, in the realistic NN
potentials are typically fitted with large \la\, (e.g \la\, ranges from
$1.3\,{\rm GeV}$~\cite{mhe87} to $2.3$ -- $2.5\,{\rm
GeV}$~\cite{ug68}). On the other hand, arguments based on resolving the
discrepancy of the Goldberger-Treiman relation~\cite{cs81} and the
discrepancy between the $pp\pi^0$ and $pn\pi^+$ couplings~\cite{th89}
suggest a much softer \ga\, with \la\, around $0.8\,{\rm GeV}$.
Furthermore, hadronic models of baryons with meson clouds like the
skyrmion typically have a rather soft form factor (i.e. \la $\sim
0.6\,{\rm GeV}$)~\cite{ll87} due to its large pion cloud and such a
small \la\, is needed for the high energy elastic $pp$
scattering~\cite{lly88}. 

In view of the large uncertainty in \ga, it is high time to study it
with a lattice QCD calculation. Since our recent calculations of the
nucleon axial and electromagnetic form factors are within $10\%$ of the
experimental results~\cite{ldd94,wdl92}, a prediction of \ga\, with a
similar accuracy should be enough to adjudicate on the controversy over
the $\pi NN$ form factor. In this letter, we extend our lattice
calculation to the proton pseudoscalar form factor for a range of light
quark masses. \ga\, is obtained by considering the pion pole dominance
in $g_P(q^2)$ when the latter is extrapolated to the quark mass which
corresponds to the physical pion mass. 

In analogy with the study of the electromagnetic and axial form
factors~\cite{ldd94,wdl92} of the nucleon, we calculate the following
two- and three-point functions for the proton 
\begin{equation} \label{twopt}
G_{pp}^{\alpha\alpha}(t,\vec{p}) = \sum_{\vec{x}}e^{-i\vec{p}\cdot
\vec{x}}  \langle 0| T (\chi^\alpha(x) \bar{\chi}^\alpha(0) |0 \rangle
\end{equation}
\begin{equation} \label{threept}
G_{pPp}^{\alpha\beta}(t_f,\vec{p},t,\vec{q})=
\sum_{\vec{x}_f,\vec{x}} e^{-i\vec{p}\cdot\vec{x}_f
  +i\vec{q}\cdot\vec{x}} \langle 0| T(\chi^\alpha(x_f) P(x)
  \bar{\chi}^\beta(0)) |0 \rangle ,
\end{equation}
where $\chi^\alpha$ is the proton interpolating field and $P(x)$ is the
mean-field improved isovector pseudoscalar current for the Wilson
fermion 
\begin{equation}   \label{pc}
P(x)= \frac{2 \kappa}{ 8\kappa_c} e^{m_q a} \bar{\psi}(x)i\gamma_5
\frac{\tau_3}{2}\psi(x).
\end{equation}
Here, we have included the $2\kappa/ 8\kappa_c$ ($\kappa_c = 0.1568$ is
the critical $\kappa$ value for the chiral limit for our lattice at
$\beta = 6.0$) and the $e^{m_q a}$ ( $m_q a = \ln (4\kappa_c/\kappa
-3)$ is the quark mass) factors in the definition of the lattice
current operator. These factors take into account the mean-field
improvement and finite quark mass correction for the Wilson
action~\cite{lm93} and have been shown to be an important improvement
in the evaluation of the axial form factor in order to allow the
perturbative lattice renormalization to work~\cite{ldd94}. 

Phenomenologically, the pseudoscalar current matrix element is written
as 
\begin{equation}
 \langle \vec{p} s| P(0) |\vec{p}^{\;\prime} s^\prime\rangle
 = g_P(q^2) \bar{u}(\vec{p}, s) i\gamma_5
 u(\vec{p}^{\;\prime},s^\prime).
\end{equation}
where $g_P(q^2)$ is the pseudoscalar form factor.
It has been shown~\cite{ldd94,wdl92} that when $t_f - t$ and $t >> a$,
the lattice spacing, the combined ratios of three-point and two-point
functions with different momentum transfers lead to desired form
factors related to the probing currents. In the case of the
pseudoscalar current in eq.~(\ref{pc}), the lattice pseudoscalar form
factor $g_P^L(q^2)$ is determined from the following ratio 
\begin{equation} \label{g4}
\frac{\Gamma^{\beta\alpha}G_{pPp}^{\alpha\beta}(t_f,\vec{0},t,
\vec{q})} {G_{pp}^{\alpha\alpha}(t_f,\vec{0})}
\frac{G_{pp}^{\alpha\alpha}(t,\vec{0})}{G_{pp}^{\alpha\alpha}
(t,\vec{q})} \longrightarrow   \frac{q_3}{E_q + m} g_P^L(q^2)
\end{equation}
where
$\Gamma=\left(\begin{array}{cc}
                 \sigma_3  & 0 \\
                 0 & 0\end{array}\right)$,
and $m$ and $E_q$ are the proton mass and energy with momentum
$\vec{q}$ respectively. 

Quark propagators have been generated on $24$ quenched gauge
configurations on a $16^3 \times 24$ lattice at $\beta = 6.0$ to study
the nucleon electromagnetic and axial form factors~\cite{ldd94,wdl92}.
We shall use the same propagators for the present calculation. Results
are obtained for three light quarks with $\kappa = 0.154$, $0.152$, and
$0.148$. They correspond to quark masses $m_q$ of about $120$, $200$,
and $370\,{\rm MeV}$ respectively. (The scale $a^{-1}=1.74(10)\,{\rm
GeV}$ is set by fixing the nucleon mass to its physical value.) Results
of $g_P^L(q^2)$ for the momentum transfers $\vec{q}\,^2 a^2 =
n(2\pi/L)^2$ ($n$ = $1$ to $4$) are obtained from the plateaus of the
ratio in eq.~(\ref{g4}) as a function of $t$, the time slice of the
current insertion, away from the sink and source of the nucleon
interpolation fields~\cite{ldd94}. Since the ratio in eq.~(\ref{g4}) is
proportional to $q_3$, $g_P^L(q^2)$ at $q^2 = 0$ can not be obtained
directly. Rather, it will be obtained from extrapolation from the
finite $q^2$ data as explained later. 

Plotted in fig.~1 are the lattice isovector pseudoscalar form factors
$g_P^L (q^2)$ of the proton as a function of the quark mass in
dimensionless unit $m_q a$ which takes into account the
tadpole-improved definition for the quark mass. They include different
momentum transfers with $\vec{q}\,^2 $ from $1$ to $4$ times
$(2\pi/La)^2$ (N.B.\ $q^2 = (E - m_N)^2 - \vec{q}\,^2$ for the
four-momentum transfer squared). The errors are obtained through the
jackknife in this case. The extrapolation of $g_P^L$ to the quark mass
$m_q a$ which corresponds to the physical pion mass is carried out with
the correlated fit to a linear dependence on the quark mass $m_qa$ for
$\kappa = 0.154$, $0.152$ and $0.148$. The data covariance matrix is
calculated with the single elimination jackknife error for $g_P^L$
which takes into account the correlation among the gauge
configurations~\cite{ldd94,gup91}. This fitting gives $\chi^2/N_{DF} =
0.005, 0.008, 0.65,$ and 1.7 for $\vec{q}\,^2 a^2$ form 1 to 4
$(2\pi/L)^2$. 

To extract the $\pi NN$ form factor \ga, we take the pion pole
dominance in the dispersion relation for $g_P^L(q^2)$ so that 
\begin{equation}  \label{gp}
g_P^L(q^2) = \frac{G_{\pi}\,\,g_{\pi NN}(q^2)}{m_{\pi}^2 - q^2}
\end{equation}
where $G_{\pi} = \langle 0|P(0)|\pi\rangle$ can be obtained from the
two-point function 
\begin{equation} \label{Gpi}
\langle 0 | \sum_{\vec{x}} P(\vec{x},t) P(0,0) | 0 \rangle _ {\,\,
\stackrel{\longrightarrow}{t\,>>\,a}}\,\,
\frac{G_{\pi}^2}{2m_{\pi}} e^{-m_{\pi}t}
\end{equation}
Plotted in Fig.~2 is \ga\, defined via eqs.~(\ref{gp}) and (\ref{Gpi}).
There is a caveat to extracting \ga\, this way which we wish to point
out. Strictly speaking eq.~(\ref{gp}) is equivalent to PCAC where the
physical pion field dominates and is thus valid for small $q^2$. For
$q^2$ as large as $m_{\pi'}^2$ with $\pi'$ being the radially excited
pion at $1.3\,{\rm GeV}$, higher mass contribution to $g_P^L(q^2)$ may
not be negligible. However, $g_{\pi' NN}$ is expected to be an order of
magnitude smaller than $g_{\pi NN}$ due to the the fact that a node in
the internal $q\bar{q}$ wavefunction of $\pi'$ will lead to
cancellation in the vertex function. Therefore, we estimate that the
pion pole dominance (eq.~(\ref{gp})) may have an error as large as 5 to
10\% at the highest $q^2$ we calculated. This is much smaller than the
statistical error we have at the highest $q^2$. This is also consistent
with the estimate that PCAC and chiral perturbation is good to a scale
of $4\pi f_{\pi}$. Keeping this in mind, we discuss the behavior of
\ga. We fitted it with both a monopole form (i.e. eq.~(\ref{pnnff}))
and a dipole form. We found that the monopole form with $\Lambda_{\pi
NN} = 0.75 \pm 0.14 \,{\rm GeV}$ and a $\chi^2/N_{DF} = 0.13/2$ is only
slightly better than the dipole form with a dipole mass of $1.32 \pm
0.17\,{\rm GeV}$ and a $\chi^2/N_{DF} = 0.57/2$, in so far as the
$\chi^2$ is concerned. However, we can inject our knowledge at $q^2=
0$. The Goldberger-Treiman (GT) relation which relates coupling
constants at $q^2 = 0$ 
\begin{equation} \label{gt}
m_N\,g_A(0) = f_{\pi} g_{\pi NN}(0)
\end{equation}
predicts $g_{\pi NN}(0) = 12.66 \pm 0.04$ from the known $g_A(0) =
1.2573 \pm 0.0028$ and $f_{\pi} = 93.15 \pm 0.11 \, {\rm
MeV}$~\cite{pdt92}. Extrapolation of the monopole fit of \ga\, gives
$g_{\pi NN}(0) = 12.2 \pm 2.3$ which agrees with the GT relation. Yet,
the dipole fit giving $g_{\pi NN}(0) = 10.8 \pm 1.3$ falls outside the
prediction of the GT relation. Thus, we conclude that the monopole form
is much preferred over the dipole form. The monopole mass $\Lambda_{\pi
NN} = 0.75 \pm 0.14\,{\rm GeV}$ thus obtained is much smaller than
those typically used in the NN potential, but agrees well with those
based on the consideration of the GT relation~\cite{cs81}, the apparent
discrepancy between $g_{\pi^0 pp}$ and $g_{\pi^+ np}$~\cite{th89}, and
the nucleon models like the skyrmion~\cite{ll87}. To salvage the nice
fit of the NN scattering data and the deuteron properties based on a
hard $\pi NN$ form factor, attempts have been made to incorporate a
soft \ga\, either by appending a heavy pion at $\sim 1.2\,{\rm
GeV}$~\cite{ht90} or by including multi-meson exchanges \cite{ued92}
(e.g. $\pi \rho$ and $\pi \sigma$). Extrapolating \ga\, to $q^2 =
m_{\pi}^2$, we obtain $g_{\pi NN}$, the $\pi N$ coupling constant, to
be $12.7 \pm 2.4$. This compares favorably with the empirical value of
$13.40 \pm 0.17$~\cite{hol83} and $13.13 \pm 0.07$~\cite{arn93}. The
4\% change in \ga\, from $q^2 = 0$ to $m_{\pi}^2$ indeed can account
for the 4\% discrepancy in the GT relation when the physical $g_{\pi
NN}$ is used in eq.~(\ref{gt}) instead of the $g_{\pi
NN}(0)$~\cite{cs81}. 

Putting the chiral Ward identity $\partial_{\mu} A_{\mu}^a = 2 m
\bar{\Psi} i \gamma_5 \tau_a/2 \Psi$ with pion pole dominance or
equivalently PCAC ($\partial_{\mu} A_{\mu}^a = f_{\pi} m_{\pi}^2
\phi^a$) between nucleon states, we find 
\begin{equation} \label{ghp}
2m_N g_A(q^2) + q^2 h_A(q^2) = \frac{2 m_{\pi}^2 f_{\pi}\,
g_{\pi NN}(q^2)} {m_{\pi}^2 - q^2}
\end{equation}
In addition to PCAC, if one further assumes that the induced
pseudoscalar form factor $h_A(q^2)$ is dominated by the pion pole, i.e.
$h_A(q^2) = 2 f_{\pi}\, g_{\pi NN}(q^2) /(m_{\pi}^2 - q^2)$, then
$g_{\pi NN}(q^2) = (m_N/f_{\pi}) g_A(q^2)$. In other words, \ga\, has
the same $q^2$ dependence as $g_A(q^2)$ which has been frequently used
in the literature~\cite{aff79,cde93}. As there is no a priori reason
why \ga\, should have the same falloff as $g_A(q^2)$ at all $q^2$ and,
furthermore, chiral perturbation calculation~\cite{gss88} at one loop
suggests that they acquire different contributions, we compare \ga\,\,
from eq.~(\ref{gp}) and $g_A(q^2)$ obtained on the same set of gauge
configurations~\cite{ldd94} for the light quark cases. Both \ga\, and
$g_A(q^2)$, normalized at $q^2=0$, are plotted in Fig. 3 for $\kappa =
0.148, 0.152, 0.154$, and 0.1567. The last $\kappa$ corresponds to the
physical pion mass. We find that in all these light quark cases, there
is a tendency for the normalized \ga\, to lie lower/higher than the
normalized $g_A(q^2)$ at lower/higher $- q^2$. This presumably reflects
the preferred monopole vs dipole fit for the \ga and $g_A(q^2)$. Our
data do  not discern this well though. If this behavior is verified, it
would imply that the induced pseudoscalar form factor $h_A(q^2)$ (not
the pseudoscalar form factor $g_P(q^2)$) is not entirely dominated by
the pion for higher $-q^2$ as it is at very low $-q^2$ ($< 0.1\,{\rm
GeV}^2$ say). 

Lastly, from the chiral Ward identity (eq.~(\ref{ghp})), we can obtain
the induced pseudoscalar form factor $h_A(q^2)$ from
$g_A(q^2)$~\cite{ldd94} and \ga\,. The pion decay constant $f_{\pi}$
needed in eq.~(\ref{ghp}) is calculated from the two-point functions
$\langle \Sigma_{\vec{x}} A_4(t, \vec{x}) P(0,0)\rangle$ and
eq.~(\ref{Gpi}). It is found to be $89.8 \pm 4.5\,{\rm MeV}$ when the
finite lattice renormalization is taken into account. We plot
$h_A(q^2)$ in Fig.~4. Also plotted in the insert are experimental data
obtained from pion electroproduction~\cite{cde93}. It turns out that
the momentum transfer ranges of our lattice calculation and the
available experiment do not overlap. We can not compare them directly.
However, if we use the monopole fit of \ga\, and the dipole fit of
$g_A(q^2)$~\cite{ldd94}, we find the extrapolation of $h_A(q^2)$ (solid
line in Fig.~4) does agree with the experimental data at small $-q^2$.
We note the errors of the fit start to diverge as $-q^2 \rightarrow 0$
due to the $q^2$ singularity in eq.~(\ref{ghp}). As a result we are not
able to extrapolate to $-q^2 = 0.88 m_{\pi}^2$ to compare with the the
muon capture experiment. 

To conclude, we have calculated the isovector pseudoscalar form factor
of the nucleon in a lattice QCD calculation for quark masses from about
one to about two times that of the strange quark. From these we
extracted \ga\, with the help of the pion pole dominance. The main
results we gleaned are the following: 

1) Incorporating the Goldberger-Treiman relation at $q^2 = 0$, we find
that \ga\, is much better described by a monopole than a dipole form.
The monopole mass \la\, $= 0.75 \pm 0.14\,{\rm GeV}$ is much smaller
than commonly used in the NN potential. 

2) $g_{\pi NN} = 12.7 \pm  2.4 $ agrees well with the phenomenological
values of $13.40 \pm 0.17$~\cite{hol83} and $13.13  \pm
0.07$~\cite{arn93}. It is also consistent with the lattice calculation
of $14.8 \pm 0.6$ with staggered fermion~\cite{agh94}. 

3) The falloff of \ga\, is about the same as $g_A(q^2)$ at very small
$-q^2$ ($< 0.3\,{\rm GeV}^2$), but is likely to fall slower at higher
$-q^2$. This suggests that the induced pseudoscalar form factor
$h_A(q^2)$ is not entirely dominated by the pion pole at higher $-q^2$.
This point needs to be verified further with higher statistics study. 

4) From the chiral Ward identity and PCAC, we obtain $h_A(q^2)$ which
can be checked experimentally in the future. 

The soft \ga\, form factor agrees with the predictions based on the
discrepancy of the Goldberger-Treiman relation~\cite{cs81} and between
the $pp \pi^0$ and $pn \pi^+$ couplings~\cite{th89}. This will have a
large impact on the study of NN potential, the three-body force, and
other processes which involve the $\pi N$ coupling. For future studies,
it is essential to improve the calculation by expanding the volume in
order to access smaller $-q^2$ and incorporating dynamical fermions
effects. 


This work is partially supported by DOE Grant No.\ DE-FG05-84ER40154
and NSF Grants Nos.\ STI-9108764 and PHY-9203306. 


%
\newpage
\noindent
{\large\bf Figure Captions}\vskip\baselineskip
\noindent
Fig.~1 \hspace{1ex} The lattice isovector pseudoscalar form factors at
various $-q^2$ as obtained from eqs.~(\ref{g4}) are plotted as a
function of $m_q a$, the quark mass in lattice unit, for the three
light quark cases (Wilson $\kappa = 0.148, 0.152$, and 0.154). The top
curve is for $\vec{q}\,^2 = (2\pi/La)^2$, the rest are for $\vec{q}\,^2
$ from $2$ to $4$ times of $(2\pi/La)^2$ in descending order. 
\vskip\baselineskip
\noindent
Fig.~2 \hspace{1ex} $g_{\pi NN}(q^2)$ at the quark mass which
corresponds to the physical pion mass. The solid and the dashed curves
represent the monopole and dipole fits with the respective monopole and
dipole mass $\Lambda$. They give quite different extrapolations at $q^2
= 0$. 
\vskip\baselineskip
\noindent
Fig.~3 \hspace{1ex} Comparison of $g_A(q^2)$ and $g_{\pi NN}(q^2)$
(both normalized to $1$ at $q^2 = 0$) as a function of $-q^2$ for the
light quark cases ($\kappa = 0.148$, $0.152$, $0.154$, and $0.1567$). 
\vskip\baselineskip
\noindent
Fig.~4 \hspace{1ex}  The induced pseudoscalar form factor $h_A(q^2)$
from eq.~(\ref{ghp}). The solid line is from the fits to $g_A(q^2)$ and
$g_{\pi NN}(q^2)$. Also plotted in the insert are data from the
electroproduction of pion~\cite{cde93}. The typical size of the error
bars for the solid line is indicated in the insert. 


\newpage

\begin{thebibliography}{99}

\bibitem{er83}
T.E.O. Ericson and M. Rosa-Clot, Nucl. Phys. {\bf A405}, 497 (1983);
Annu. Rev. Nucl. Part. Sci. {\bf 35}, 271 (1985).
\bibitem{mhe87}
R. Machleidt, K. Holinde, and Ch. Elster, Phys. Rep. {\bf 149},
1 (1987).
\bibitem{ug68}
T. Ueda and A.E.S. Green, Phys. Rev. {\bf 174}, 1304 (1968).
\bibitem{cs81}
S. Coon and M.D. Scadron, Phys. Rev. {\bf C23}, 1150 (1981);
$\pi N$ Newsletter {\bf 3}, 90 (1991).
\bibitem{th89}
A. Thomas and K. Holinde, Phys. Rev. Lett. {\bf 63}, 2025 (1989).
\bibitem{ll87}
B. A. Li and K.F. Liu, \underline{Chiral Solitons}, pp. 421, ed. K.F.
Liu (World Scientific, 1987).
\bibitem{lly88}
B. A. Li, K. F. Liu, and M. L. Yan, Phys. Lett. {\bf 212B}, 108 (1988).
\bibitem{ldd94}
K.F. Liu, S. J. Dong, T. Draper, J. M. Wu, and W. Wilcox,
Phys. Rev. {\bf D49}, 4755 (1994).
\bibitem{wdl92}
W. Wilcox, T. Draper, and K.F. Liu, Phys. Rev. {\bf D46}, 1109 (1992).
\bibitem{lm93}
G. P. Lepage and P.B. Mackenzie, Phys. Rev. {\bf D48}, 2250 (1993).
\bibitem{gup91}
R. Gupta et al., Phys. Rev. {\bf D44}, 3272 (1991).
\bibitem{pdt92}
Review of Particle Properties, Phys. Rev. {\bf D45} (1992).
\bibitem{ht90}
K. Holinde and A. W. Thomas, Phys. Rev. {\bf C42}, R1195 (1990).
\bibitem{ued92}
T. Ueda, Phys. Rev. Lett. {\bf 68}, 142 (1992);
G. Janssen, J. W. Durso, K. Holinde, B. C. Pearce, and J. Speth,
Phys. Rev. Lett. {\bf 71}, 1978 (1993).
\bibitem{hol83}
G. H\"{o}hler, {\it $\pi N$ Scattering}, in Landolt-B\"{o}rnstein,
vol. 9b2, ed. H. Schopper (Springer, 1983).
\bibitem{arn93}
R.A. Arndt et al., $\pi N$ Newsletter {\bf 8}, 37 (1993).
\bibitem{aff79}
E. Amaldi, S. Fubini, and G. Furlan, {\it Pion Electroproduction},
(Springer, Berlin, 1979).
\bibitem{cde93}
S. Choi et al., Phys. Rev. Lett. {\bf 71}, 3927 (1994).
\bibitem{gss88}
J. Gasser, M.E. Sainio, and A. \u{S}varc, Nucl. Phys. {\bf B307},
779 (1988).
\bibitem{agh94}
R.L. Altmeyer, M. G\"{o}ckler, R. Horsley, E. Laerman, G. Schierholz,
and P.M. Zerwas, hep-lat/9401001.

\end{thebibliography}
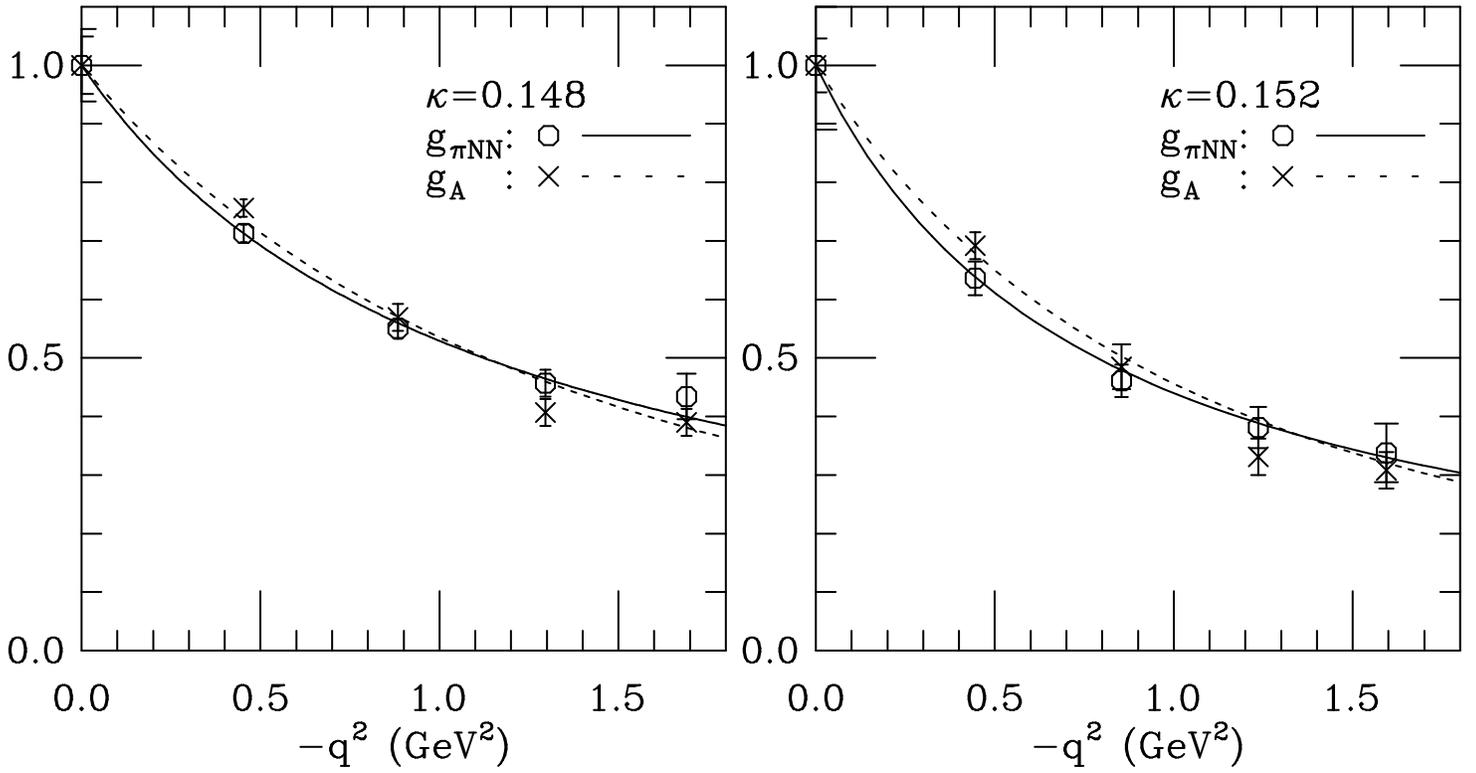
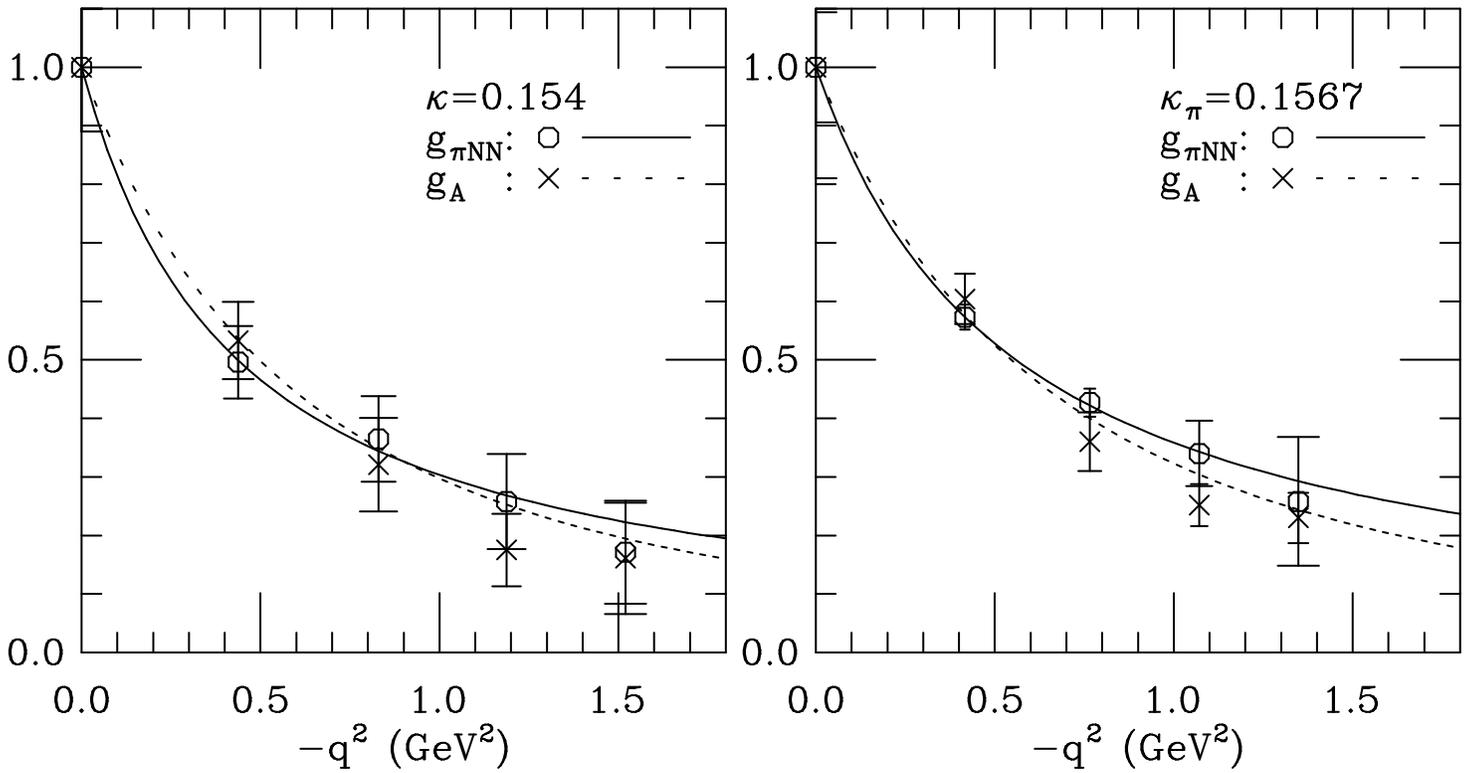
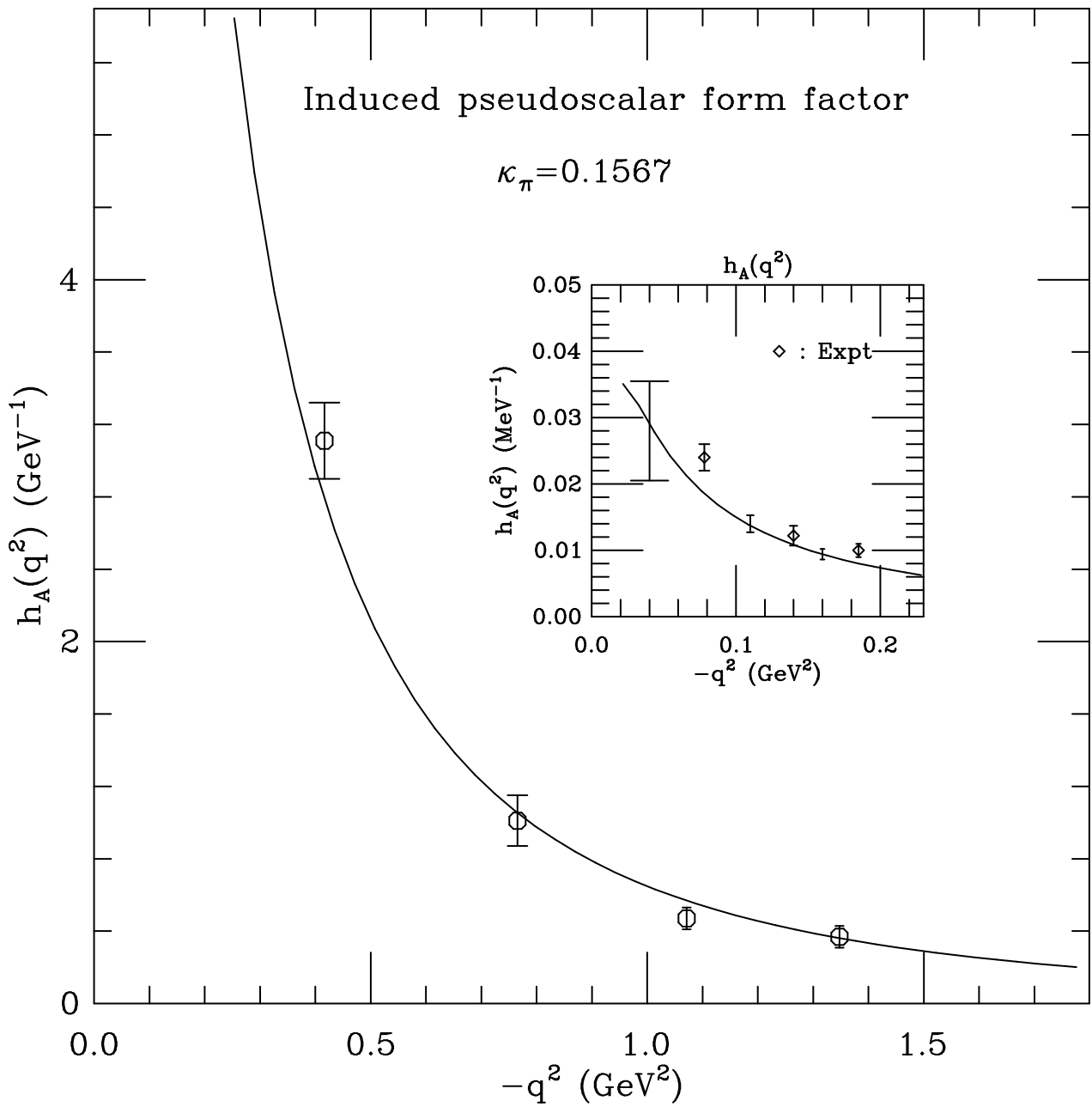
\end{document}